\documentclass[oldversion]{aa}
\usepackage{graphicx}
\usepackage{times}
\usepackage{natbib}

\newcommand{\kms}{\,km\,s$^{-1}$~}	  
\newcommand{\lya}{Ly$\alpha$}

\newcommand{\esca}{erg\,s$^{-1}$\,cm$^{-2}$\,arcsec$^{-2}$~}
\newcommand{\escan}{erg\,s$^{-1}$\,cm$^{-2}$\,\AA$^{-1}$~}

\newcommand{\br}{BR~1033$-$0327}
\newcommand{\q}{Q~2139$-$4324}
\newcommand{\sdss}{SDSS~J0939$+$0039}
\def\kmsmpc{km s$^{-1}$ Mpc$^{-1}$}

\begin{document}
\title{Deep optical spectroscopy of extended Ly$\alpha$ emission \\
around three radio-quiet $z=4.5$ quasars
\thanks{Based on
observations made with the  FORS2 multi-object spectrograph mounted on
the Antu  VLT   telescope  at   ESO-Paranal   Observatory   (programme 
079.B-0132B; PI: P. North)}}
\author{F.~Courbin\inst{1} \and P.~North\inst{1} \and A.~Eigenbrod\inst{1}
\and D.~Chelouche\inst{2,}\thanks{Chandra Fellow}}
%
%
\institute{Laboratoire d'Astrophysique, 
           Ecole Polytechnique F\'ed\'erale de Lausanne (EPFL),
           Observatoire de Sauverny,
           CH-1290 Versoix, Switzerland
\and
           Institute for Advanced Study, Einstein Drive, 
           Princeton, NJ 08540, USA}
\date{Received xxxxxx/ Accepted xxxxxx}
\authorrunning{Courbin et al.}
\titlerunning{Extended Ly$\alpha$ envelopes around three $z=4.5$ quasars}
\abstract{We 
report  the first   results   of a spectroscopic    search  for \lya\,
envelopes   around      three $z\sim4.5$  radio-quiet    quasars.  Our
observational strategy uses the FORS2 spectrograph attached to the UT1
of the Very Large Telescope (VLT) in the multi-slit mode.  This allows
us to observe  simultaneously the quasars and  several PSF stars.  The
spectra of the latter  are used to  remove the point-like  quasar from
the  data, and  to  unveil  the  faint  underlying   \lya\,  envelopes
associated  with the  quasars  with  unprecedented depth.   We clearly
detect  an envelope around two  of the three quasars.  These envelopes
measure respectively 10\arcsec\ and 13\arcsec\ in  extent (i.e. 67 kpc
and 87 kpc). This is 5 to 10 times larger than predicted by the models
of Haiman \& Rees (2001) and up to 100 times fainter. Our observations
better  agree  with models involing  a  clumpy envelope as  in Alam \&
Miralda-Escud\'e (2002) or Chelouche et al.  (2008).  We find that the
brighter quasars also have the brighter envelopes  but that the extend
of the envelopes  does not depend  on the quasar luminosity.  Although
our results are based on only two objects with a detected
\lya\, envelope, the quality of the spatial  deblending of the spectra
lends  considerable  hope  to estimate   the  luminosity function  and
surface brightness profiles of high redshift
\lya\, envelopes down to $F\sim 2-3 \times 10^{-21}$ \escan. 
We  find  that the best strategy   to carry out  such  a project is to
obtain both narrow-band images and deep slit-spectra.}


\maketitle

\section{Introduction}

Large-scale \lya\, emission   ($\sim$10$-$100  kpc) is common  among  high
redshift ($z >  2$)  radio galaxies  \citep{VM07}.  This extended  gas
shows two components: one  is kinematically perturbed  by the jets and
is part  of  a jet-induced  outflow; the  other has   a lower velocity
dispersion (a  few hundreds of \kms instead   of about $1000$~\kms) as
well as a fainter surface brightness, but may extend beyond $100$~kpc.
It has been shown, at least in the case of MRC  2104-242, that the gas
is infalling towards   the  galaxy center \citep{VM07}.  This   author
makes  the interesting suggestion that ``the  radio activity is fed by
the infalling gas, so that  it  is only detected   when the infall  is
happening and efficiently feeding the active nucleus''.

\begin{table}
\caption{Journal of observations, along with the main characteristics
of  the  quasars.  The apparent    magnitudes  are  given  in the   AB
system. They are computed by integrating the quasar's spectrum through
the {\tt  RSPECIAL} ESO filter  curve.  The absolute magnitude assumes
$H_0=72$ \kmsmpc and $(\Omega_m ,\Omega_{\lambda})=(0.3,0.7)$.}
\begin{center}
\begin{tabular}{ccccc}
\hline
redshift&JD(start)&Exposure&Airmass& Seeing	   \\
z	 &$-2400000$ &time (s)&	    & (")	   \\ \hline
\multicolumn{5}{l}{SDSS\,J09395+0039, z=4.490, R(AB)=20.9, M$_R$=-27.2} \\
\hline 
      &54203.598& 1210   & 1.18 &$1.18$\\
      &54203.612& 1210   & 1.23 &$1.24$\\
      &54203.631& 1210   & 1.31 &$1.08$\\
      &54203.645& 1210   & 1.40 &$0.84$\\
      &54203.672& 1210   & 1.64 &$0.86$\\
      &54203.686& 1210   & 1.83 &$0.81$\\ \hline
\multicolumn{5}{l}{BR\,1033-0327, z=4.509, R(AB)=18.5, M$_R$=-29.6} \\
\hline
      &54208.603& 1300   & 1.10 &$0.88$\\
      &54208.618& 1300   & 1.13 &$1.23$\\
      &54236.656& 1300   & 1.17 &$0.59$\\
      &54236.671& 1300   & 1.22 &$0.47$\\ \hline
\multicolumn{5}{l}{Q\,2139-4324, z=4.460, R(AB)=21.8, M$_R$=-26.2} \\
\hline
      &54326.566& 1300   & 1.41 &$1.09$\\
      &54326.581& 1300   & 1.32 &$0.98$\\
      &54297.632& 1300   & 1.51 &$1.61$\\
      &54297.647& 1300   & 1.40 &$1.55$\\
      &54299.631& 1300   & 1.47 &$1.29$\\
      &54299.647& 1300   & 1.37 &$1.28$\\
      &54319.669& 1300   & 1.10 &$1.47$\\
      &54319.685& 1300   & 1.08 &$1.48$\\ \hline
\end{tabular}
\end{center}
\label{basics}
\end{table}

Similar \lya\, envelopes (sometimes  called  ``blobs'') are  also  found
around radio-quiet quasars
\citep[][hereafter 
CJW]{SAS00,vB06,BSS03,WMF05,CJW06}. According to the latter authors,
\lya\, envelopes  around radio-quiet quasars (RQQs)  
are  an   order of magnitude less   luminous  than those  around RLQs,
presumably because the  emission of RLQ  gaseous envelopes is enhanced
by interactions with the radio jets.

\citet{HR01} have envisaged \lya\, emission as a possible constraint on
galaxy formation. Gas infalling into  the gravitational well of a dark
matter halo would be heated,  and dissipate this thermal energy partly
by collisional    excitation of the  \lya\,   transition.  But  the \lya\,
emission would remain too faint   to be detectable at high   redshift.
However, if  a central quasar turns  on and photoionizes the  gas, the
\lya\, emission  will  be much enhanced  and  should be detectable  with
present-day instruments.  For  quasars   at redshift   $z=3-8$,  these
authors predict surface  brightnesses  in the range $10^{-18}<   \mu <
10^{-16}$ \esca and angular  sizes of a few arcseconds.  \citet{AME02}
have  proposed another   theoretical  prediction of   the \lya\, surface
brightness  of a    quasar  host galaxy  at  $z=3$.    They find  $\mu
=10^{-17.5}$ and  $\mu = 10^{-19.5}$ \esca   at angular separations of
$0.5"$ and  $3"$  respectively from the  quasar.    This is  much more
pessimistic than the   estimate by \citet{HR01}, because  the  authors
assume a  smaller clumping factor of the  cold gas and a \lya\, emitting
region  much smaller    than   the  virial radius.  More     recently,
\citet{CMBG08} suggested that quasar nebulae are another manifestation
of metal absorption systems  associated with $L^\star$ galaxies which,
by virtue of a nearby quasar, become more efficient \lya\, emitter that
also  scatter \lya\, photons from the  broad line region  of the quasar
(BLR). The extent and luminosity of the halo in  their model serves as
a  means to study  the nearby  environments of  quasars and weigh  the
gaseous  content   of their  halos.  To summarize,    the study of the
emission properties of quasar envelopes provides a promising new means
for testing galaxy  formation  models and  various scenarios  for  the
enrichment of the inter-galactic medium.

To explore in more detail the  spatial extent, the luminosity
and kinematics of the large  hydrogen envelopes of remote quasars,
we have selected a sample of quasars at redshift $z\sim 4.5$, spanning
a  broad magnitude  range. Before carrying  a systematic  study of the
\lya\,  envelopes of these quasars, we  have obtained  deep VLT optical
spectra for  3  of them,  spanning 3  magnitudes.  The   present paper
describes our observational strategy and  our  main results, with  the
detection of a \lya\, ``nebula'' for two of the three quasars.

\begin{figure}[t!]
\centering
\includegraphics[width=9cm]{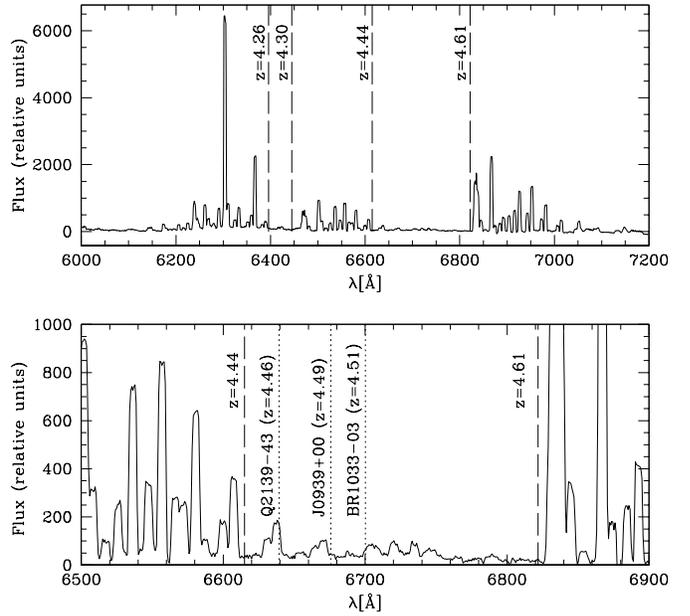}
 \caption{Spectrum of the sky emission, extracted  from our FORS2 data
 and  grism G1200R in   combination  with a 2\arcsec-slit.  The   full
 wavelength range  accessible to this  grism is displayed in the upper
 panel.  A zoom is displayed in the bottom  panel, of the region where
 falls the \lya\  emission redshifted at  z$\sim$4.5. The positions of
 the \lya\ line of the three quasars are also indicated.}  \label{sky}
\end{figure}  

\begin{figure}[t!]
\centering
\includegraphics[width=8.7cm]{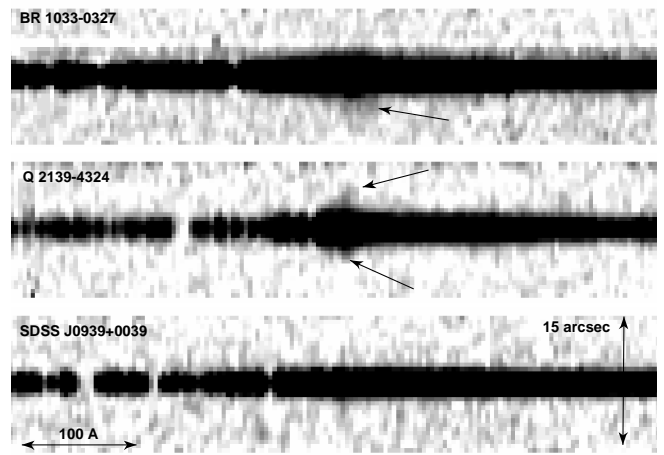}
\caption{The  combined and   sky-subtracted FORS2  spectra for the three 
quasars, prior  to any deconvolution (see  text). The images have been
binned by 5 pixels in the spectral direction (i.e., the new pixel size
is 3.8 \AA) and by  2 pixels in the spatial  direction, resulting in a
pixel  size of 0.5\arcsec.   The extended  \lya\ emission is indicated
with arrows in two of the objects. }
\label{binned_spectra}
\end{figure}

\begin{figure*}[t!]
\centering
\includegraphics[width=12cm]{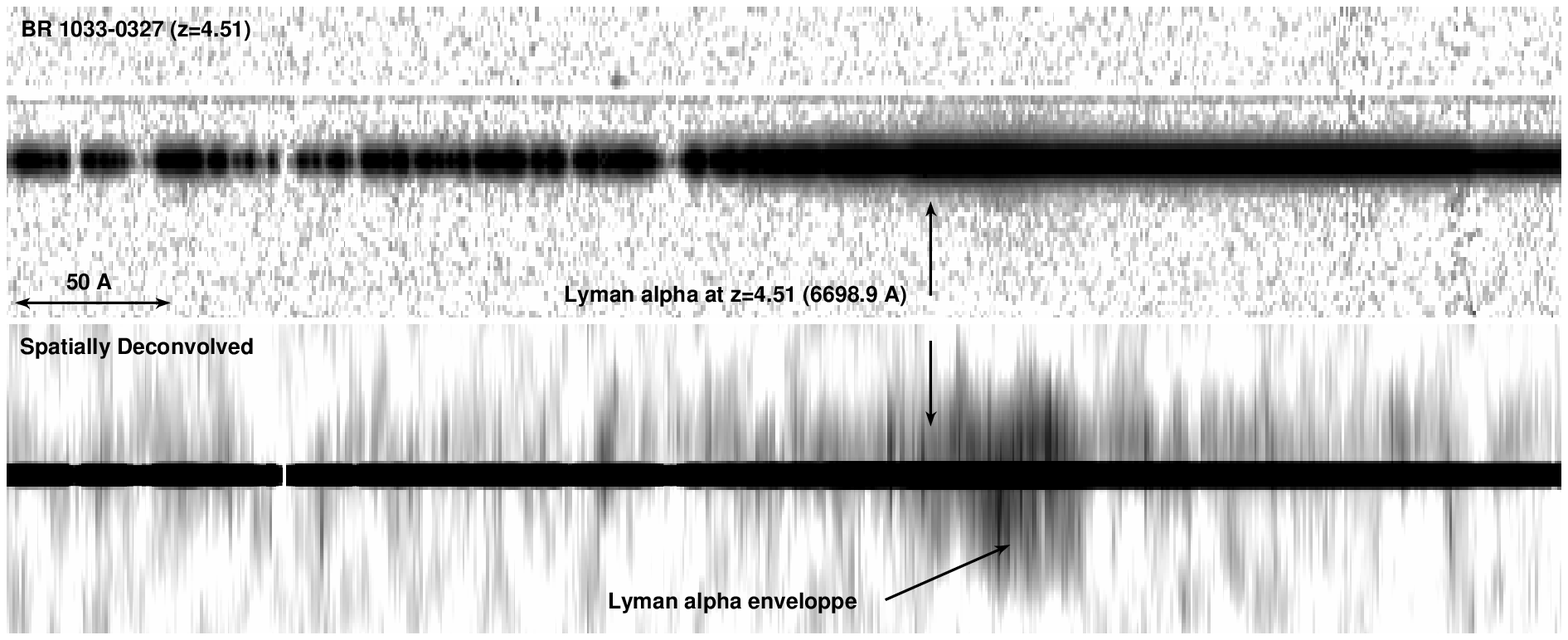}\vskip 10pt
\includegraphics[width=12cm]{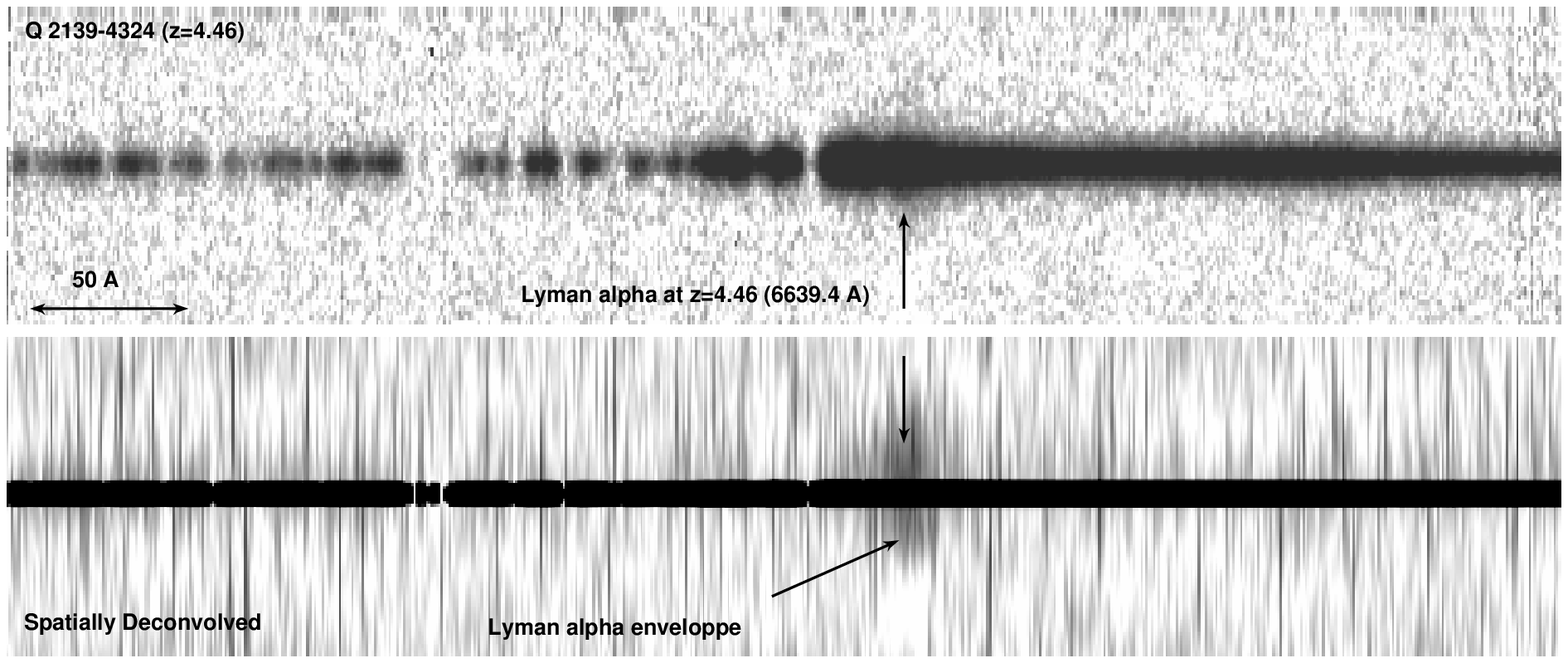}\vskip 10pt
\includegraphics[width=12cm]{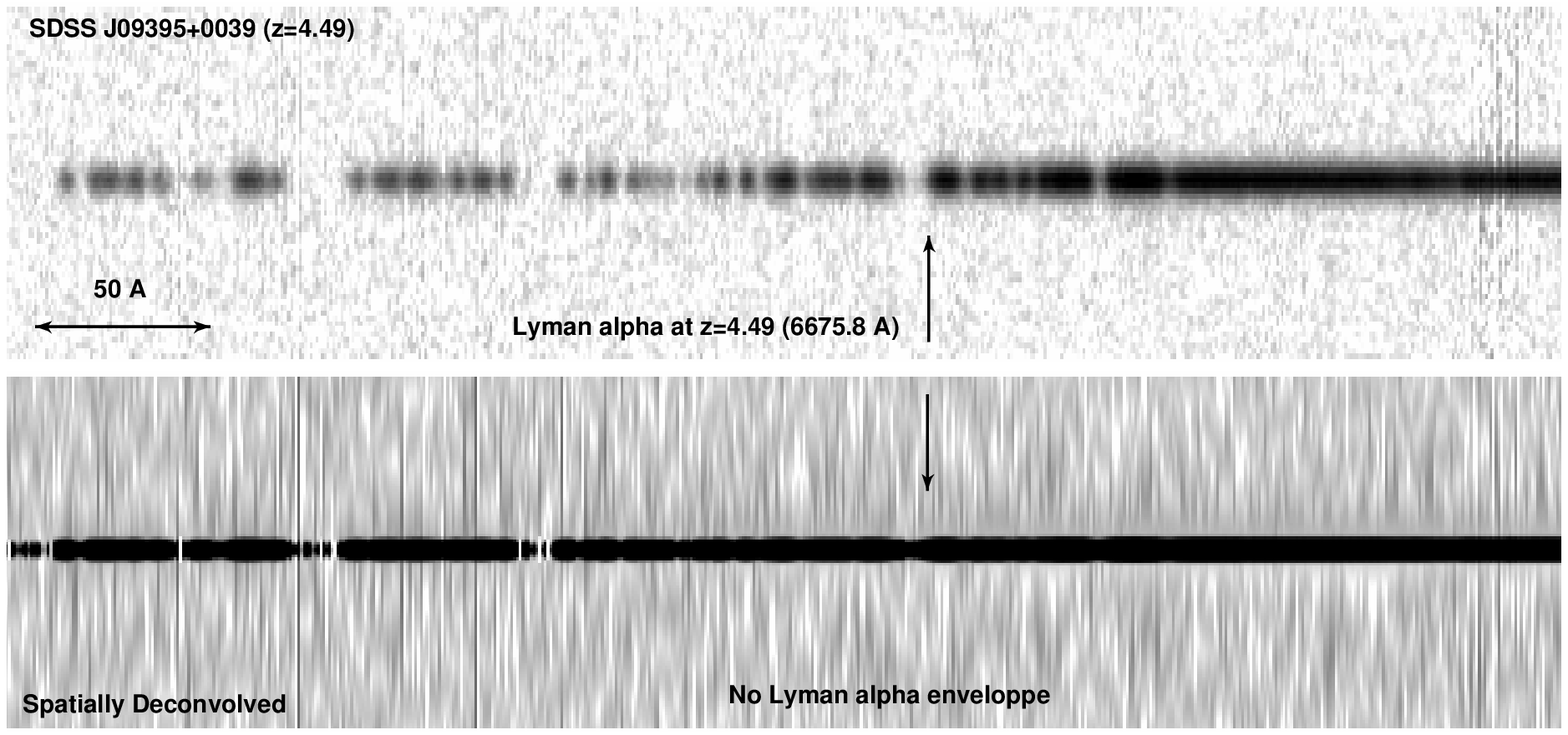}
\caption{The three combined and sky-subtracted FORS2 spectra along
with their  spatial deconvolution.     The spatial  resolution   after
deconvolution  is  0.25\arcsec.  In   each  panel, the  vertical arrow
indicates  the  position of  the  \lya\  emission  line of the
quasar.  A \lya\  envelope is detected  in two  quasars out of
three. The envelope of \br\ is redshifted compared with the 
quasar. The height of the spectra is 16\arcsec\ in all panels.}
\label{2Dspectra}
\end{figure*}

\section{Deep VLT optical spectroscopy}

Motivated by the hypothesis that the \lya\, emission around quasars is
enhanced  by the quasar  radiation field, our  goal is to characterize
the  properties  of the  \lya\,  envelopes  as function of  the quasar
luminosity. The construction of a sample is  a compromise between this
scientific goal and the technical constraints imposed by the intrinsic
faintness of the \lya\, envelopes. The observations presented here are
intended to demonstrate the  feasibility of the project  and to give a
first characterization of the \lya\, envelopes  for three quasars with
different luminosities.

\subsection{Sample and observations}

We select bright quasars from the 3rd edition of the SDSS catalogue of
quasars \citep{SCHNEIDER05} and from the 12th edition of the V\'erons'
catalogue \citep{VV06}.  All our targets are at high redshift,
i.e., when  galaxy formation  was still in  place.
\sdss\, and \q\, 
have not been explicitly  related to any radio source \citep{VV06}  but
given their  brightness, they   are  likely RQQs.    \br\, is  an  RQQ
according to  \cite{VV06} and \cite{B03}.  

A  key observational  issue is  to avoid  both the telluric absorption
lines and the numerous  atmospheric emission lines in  the red part of
the  optical domain.  Residuals of   such lines in the  sky-subtracted
spectra may   mimic a  \lya\,    envelope.  This  latter  consideration
restricts the accessible redshift range to $4.44 < z < 4.61$, as shown
in   Fig.~\ref{sky}, with an  additional small  window at  $4.26 < z <
4.30$.  Finally,  there must be  adequate nearby stars  with about the
same  brightness  as   the quasar  in   order   to carry out  a careful
subtraction of  the   quasar's light  using  image  deconvolution,  as
described in \citet{COU2000}.

\begin{figure*}[t!]
\centering
\includegraphics[width=8.8cm]{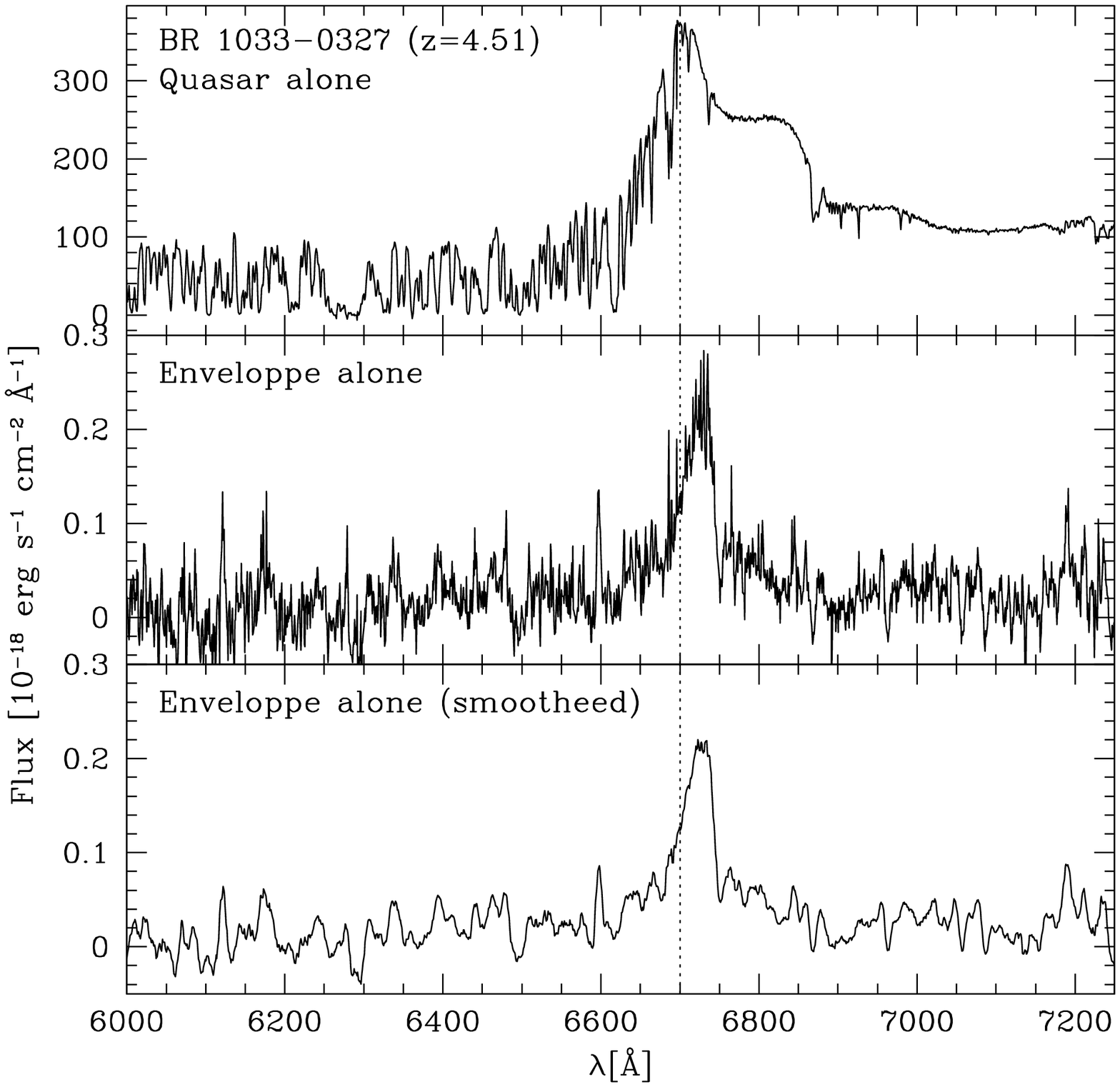}
\includegraphics[width=8.8cm]{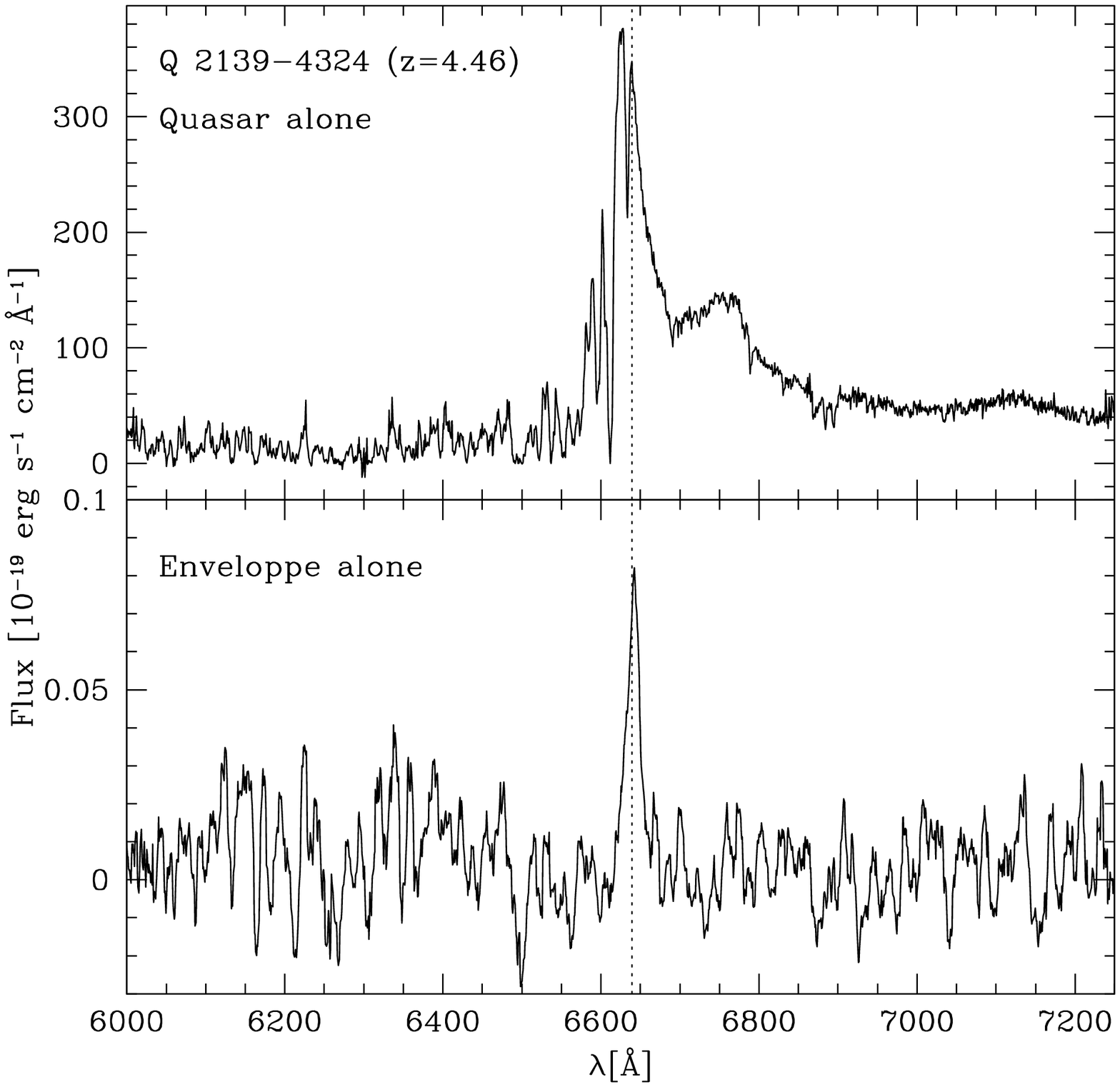}\\
\includegraphics[width=8.8cm]{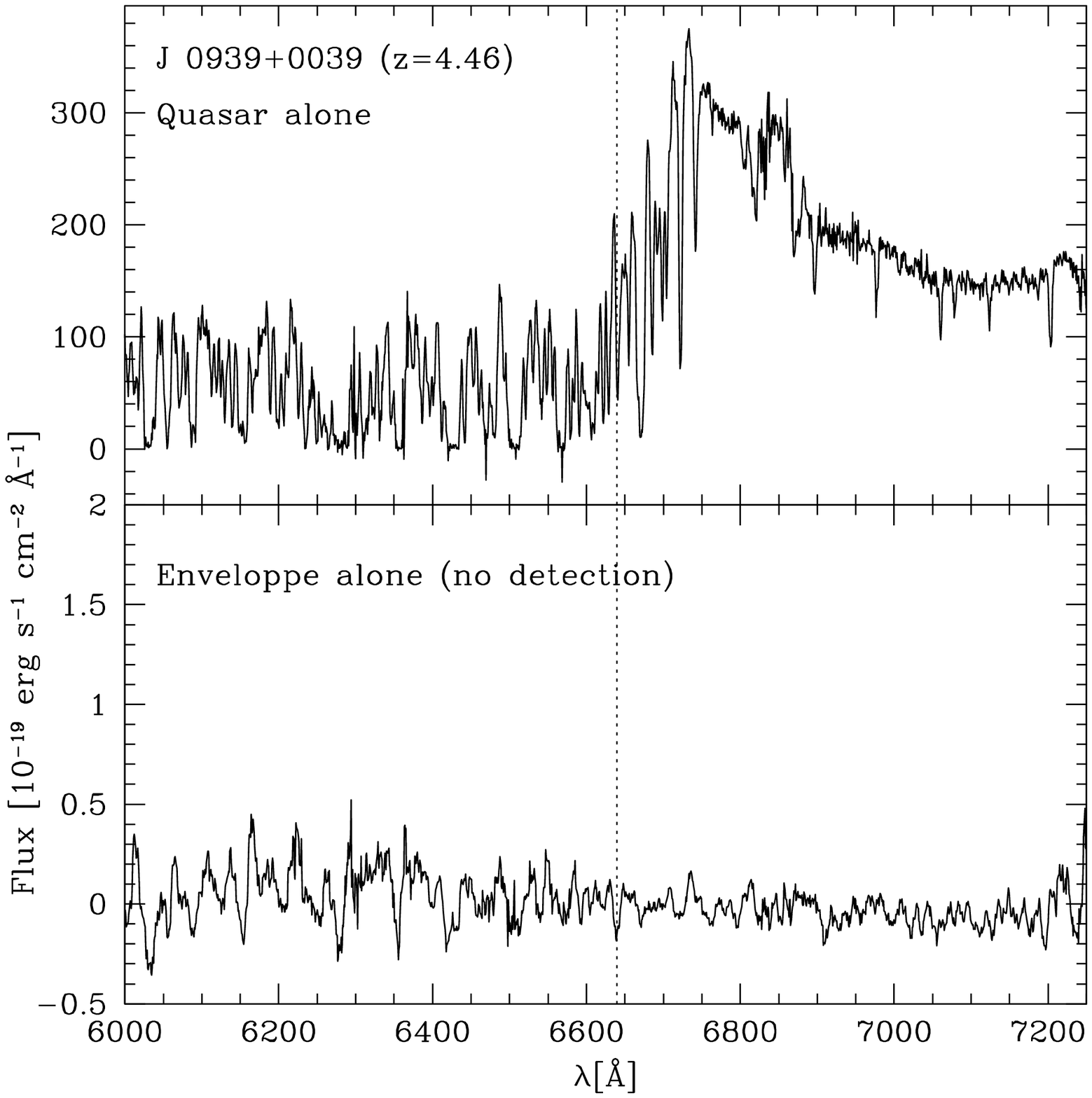}
 \caption{Extracted 1D spectra  for the three  quasars.  In each case,
 the top  panel shows the quasar alone  and the bottom panel shows the
 \lya\, envelope  alone, after  spatial  deconvolution of the  spectra.
 The vertical dotted  line  indicates the position of  the un-absorbed
 \lya\, emission line, at the redshift of the quasar.  In
 all cases, the spectrum of the \lya\, envelope has been smoothed using
 a  boxcar of   8  \AA. In  the   case of  \br\  we also  display the
 un-smoothed spectrum (middle  panel), as  the  
 signal-to-noise is higher than in the other two objects.} 
\label{1Dspectra}
\end{figure*}



The  observations were carried  out  in ESO Period 79 (April-September
2007)  in   service  mode  with the    FORS2 multi-object spectrograph
attached to  VLT-UT1.  The ESO  grism G1200R+93 has a  resolving power
$R=1070$ with a  2\arcsec-slit, ensuring to  catch most of the flux of
the \lya\, envelope.   This  grism is  used  in  combination with  the
GG435+81 order separating  filter, leading to the wavelength  coverage
6000 \AA\  $< \lambda  <$  7200 \AA.   The maximum efficiency  of this
combination coincides well  with   the  expected wavelength of     the
redshifted \lya\ line, i.e. about 6686 \AA.

The multi-slit  MXU mode  is used,  with  slits that are   long enough
(typical  length: $\sim 20"$) to reliably  model and subtract the sky
emission.  Although  only one scientific target  is observable in each
field,  we use the  MXU capability  in order  to observe several stars
through identical  slits.  In  this way,  a  spectrum PSF is  measured
simultaneously with  the  quasar.  This  is  crucial  for the  spatial
deconvolution of the data  to work efficiently \citep{COU2000} and  to
separate well the quasar spectrum from that of the putative envelope.

Although the requested maximum seeing is 0.9\arcsec\ for our observing
programme, this  condition is fulfilled in  practice for three Observing
Blocks (OBs) out of the nine executed. Each planned OB consists of one
short    acquisition  image, followed   by  two    long  exposures  for
spectroscopy.  The planned total exposure time  for each object is 2.7
hours, with seeing better than 0.9\arcsec.  The  journal of the actual
observations is presented in Table~1.


\begin{table*}[t!]
\caption{Main properties of the \lya\, envelopes, as measured on the 
spectra of  Fig.~\ref{1Dspectra}.   All parameters  are given  in  the
observer's frame.  The
$1-\sigma$ detection limit is the standard deviation of the background
noise (after smoothing with a car box  of 8~\AA ) integrated along the
whole slit.  Note  that  the   surface  brightness is integrated    in
wavelength but  given  per arcsec$^2$, while   the $1-\sigma$ limit is
spatially integrated but given per \AA.}
\begin{center}
\begin{tabular}{lcccccc}
\hline
Object & $\lambda$ & mean F$_{\lambda}$              & FWHM  & Extent & Surface brightness &	$1-\sigma$ detection limit 	   \\
       & [\AA] &  [erg s$^{-1}$ cm$^{-2}$ \AA$^{-1}$]  & [\AA]   & (\arcsec, kpc)  & [erg s$^{-1}$ cm$^{-2}$ \arcsec$^{-2}$] 	     &    [erg s$^{-1}$ cm$^{-2}$ \AA$^{-1}$]       \\ 
\hline
\br      & $6725.0 \pm 0.5$  & $4.0(\pm 0.4)\times 10^{-19}$ & $50 \pm 10$  & 13, 86  & $7.7(\pm 0.8)\times 10^{-19}$ & $2.7 \times 10^{-20}$\\
\q       & $6641.0 \pm 0.3$  & $7.2(\pm 1.4)\times 10^{-21}$ & $22 \pm 2$   & 10, 66 & $8.0(\pm 1.6) \times 10^{-21}$ & $2.5\times 10^{-21}$\\  
\sdss    &         $-$       &         $-$          &     $-$      & $-$ &         $-$          & $2.0\times 10^{-20}$\\
\hline
\end{tabular}
\end{center}
\label{lyman}
\end{table*}

\subsection{Reduction and spatial deconvolution}

The data  reduction is straightforward.  It is   carried out using the
standard  {\tt  IRAF} procedures.  The   individual  spectra listed in
Table~1   are   flatfielded     using   dome    flats,    and     then
wavelength-calibrated in two dimensions  in  order to correct the  sky
emission lines for  slit curvature. The  scale of the reduced  data is
0.76  \AA\, per pixel in  the  spectral direction and 0.25\arcsec\, in
the spatial direction.

The   sky emission is  then  subtracted from the  individual frames by
fitting a  second order polynomial along  the spatial direction.  This
fit considers only the 10 pixels on each side of  the slit and the sky
at the position of the quasars is interpolated using this fit, both on
the quasar and on the PSF stars.

The   cosmic rays are   removed using  the {\tt L.A.Cosmic}  algorithm
\citep{vanDokkum2001}. All the  frames are checked visually after this
process in  order to make  sure that  no  signal is mistakenly removed
from the data.   Particular care  is paid  to the good  seeing frames,
where the central parts of the quasar and of the PSF stars must not be
misidentified with cosmic rays.

Even though  FORS2 has an atmospheric  refraction corrector, the shape
of the spectra along  the spectral direction shows slight distortions,
i.e., the position of the spectrum changes
as a function of wavelength.  These  distortions are corrected for and
the spectra are eventually weighted so that  their flux is the same at
a  reference wavelength,  before they  are   combined into a deep   2D
spectrum. We show in Fig.~\ref{binned_spectra} the combined spectra of
the three quasars, after binning both in  the spatial and the spectral
directions.  An extended \lya\,  envelope  is   already visible up   to
4\arcsec\ away from the quasars in two objects: \br\ and \q.

While a  \lya\, envelope   is visible on    the combined data  already,
measuring  its actual flux  requires  accurate spatial deblending.  We
carry out this  delicate  task by spatially deconvolving  the  spectra
following the method described in \citet{COU2000}.   This method is an
adaptation to   spectroscopy  of  the  ``MCS''  image    deconvolution
algorithm \citep{MCS}.   It   has  been  used  successfully   on  many
occasions.   Among others,  the algorithm has  been  used in  order to
unveil the spectrum of the lensing galaxy in multiply imaged quasars
\citep{eigen07} and to study low redshift quasar host galaxies \citep{LMC06}. 

The results of  the present application to  high  redshift quasars are
displayed  in  Fig.~\ref{2Dspectra}.  The  algorithm uses  the spatial
information contained in the spectrum of several PSF stars in order to
sharpen the data in the  spatial direction. At  the same time, it also
decomposes the  data  into   a point-source and     an extended-source
channel.  The output of  the deconvolution  procedure consists of  two
individual spectra, one for the quasar and one for its host galaxy (or
\lya\,   envelope),   free of any  mutual  light   contamination. It is
therefore  possible to estimate the luminosity  of  the \lya\, emission
``underneath''  the quasar. Subsampling of   the data is also possible
with the  MCS algorithm, hence  the pixel size in Fig.~\ref{2Dspectra}
is half that of  the   original data,  i.e.,  the  new pixel size   is
0.125\arcsec.

\begin{table}[t!]
\caption{\lya\, luminosity of the quasar in the BLR, 
compared with the lyminosity of  the extended envelope.  The flux  in
the  quasar BLR  is  measured in the  wavelength interval 1200-1230\AA\
(rest-frame). The last two lines give the \lya\,  flux of the envelope,
after correction by slit-clipping (see text).}
\begin{center}
\begin{tabular}{lcc}
\hline
Object   &     L(BLR)         &      L(\lya)          \\
         & [erg s$^{-1}$]     &  [erg s$^{-1}$]       \\
\hline
\br      & $7.2(\pm 0.4) \times 10^{45}$ & $4.0(\pm 0.4) \times 10^{42}$  \\
\q       & $4.6(\pm 0.2) \times 10^{44}$ & $3.0(\pm 0.6) \times 10^{40}$  \\
\sdss    & $4.1(\pm 0.2) \times 10^{44}$ &         $-$           \\
\hline 
\br      &         $-$        & $2.0(\pm 0.2)\times 10^{43}$  \\
\q       &         $-$        & $1.2(\pm 0.3)\times 10^{41}$  \\
\hline
\end{tabular}
\end{center}
\label{lyman_lum}
\end{table}

\section{Results}

The  main scientific information in our  data is the \lya\, luminosity
of the   envelopes,  their  angular size,   and   their  mean  surface
brightness.    The     measurements   of  the    spectra     shown  in
Fig.~\ref{1Dspectra}     are  presented     in  Tables~\ref{lyman}  \&
\ref{lyman_lum}.    To check the flux   calibration of the deconvolved
spectra, we integrate  them in the FORS2  {\tt RSPECIAL} filter.  This
filter is also used  to obtain short  acquisition images prior  to the
long  spectroscopic exposures.   These ``spectroscopic'' AB magnitudes
are given  in  Table~1.  We check  that  they are  compatible with the
simple aperture photometry obtained from the short images.

We detect a \lya\, envelope  in 2 out of 3   objects.  Our flux  limit
(integrated over the  whole  object) is  very  faint, as indicated  in
Table~\ref{lyman}, up to two orders of magnitude fainter than previous
studies in this field \citep{CJW06, bremer92}. Note, however, that our
limit is computed by integrating the spectrum over  the full extent of
the \lya\, envelope. 

The measurable extent of the envelopes is $r\sim  43$~kpc for \br\ and
$r\sim  33$~kpc for  \q, measured from  the  quasar's centroid  to the
faintest visible isophote, i.e., at   the detection limit as given  in
Table~\ref{lyman}.

While the surface brightness  of  the \lya\,  fuzz is not  affected by
slit losses, the    total luminosity  is.  Assuming   that the  \lya\,
envelopes are  uniform face-on   disks with   diameters equal to   the
extents  quoted  in Table~\ref{lyman},  we  can estimate the amount of
flux missed by   using   a slit   width  of 2\arcsec.    The  observed
luminosities of the  \lya\, envelopes for   \br\ and \q\ are  given in
Table~\ref{lyman_lum} as well as  the luminosities after the  correction
for the slit clipping.

The velocity of the \lya\, envelope in \q\ is well compatible with that
of  the   quasar.  However, the  extended  \lya\,  emission in \br\  is
redshifted by $\Delta \lambda = 26\pm5$~\AA, i.e., $\Delta V=+1165 \pm
225$ \kms\  with respect to that of  the quasar.  We have  checked our
deconvolution using several  PSF  stars and using  different smoothing
terms   \citep[see][]{COU2000}. These checks   leave  the observed shift
unchanged.  An   explanation for this  line   shift  may be  that  the
redshift of the  quasar is incorrect, as  it is measured only from the
\lya\,  line   itself and from  the   \ion{C}{IV} and \ion{C}{III]}  
lines.  The \lya\, emission of the  quasar is strongly affected by the
\lya\,  forest,  and the  carbon lines   in quasars are  known  to  be
blueshifted with respect to the actual redshift of  the quasar.  Using
infrared observations, where the quasar  \ion{[O}{III]} narrow line is
accessible  at moderate  redshifts, \citet{Mintosh99} find  an average
blueshift   of  860\kms\, and  600\kms\,      of the \ion{C}{IV}   and
\ion{C}{III]} broad lines with respect  to the \ion{[O}{III]} line.  A
similar conclusion is drawn   by \citet{richards02} using  SDSS quasar
spectra.   The velocity shift we  observe  between the  quasar and the
\lya\, envelope of   \br, although large, is  still  compatible with a
biased redshift measurement of the quasar.



\begin{figure}[t!]
\centering
\includegraphics[width=8.7cm]{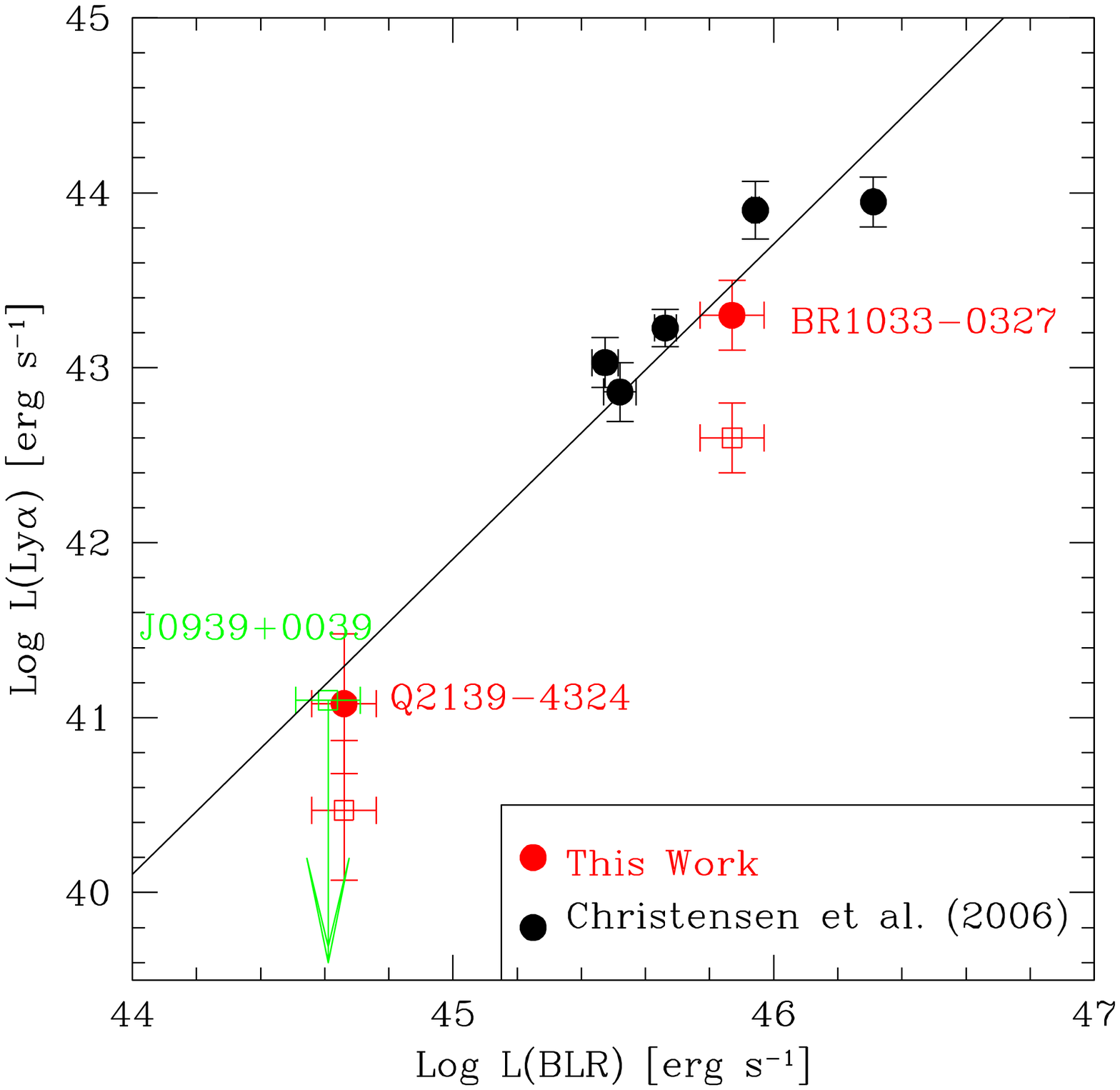}
\includegraphics[width=8.7cm]{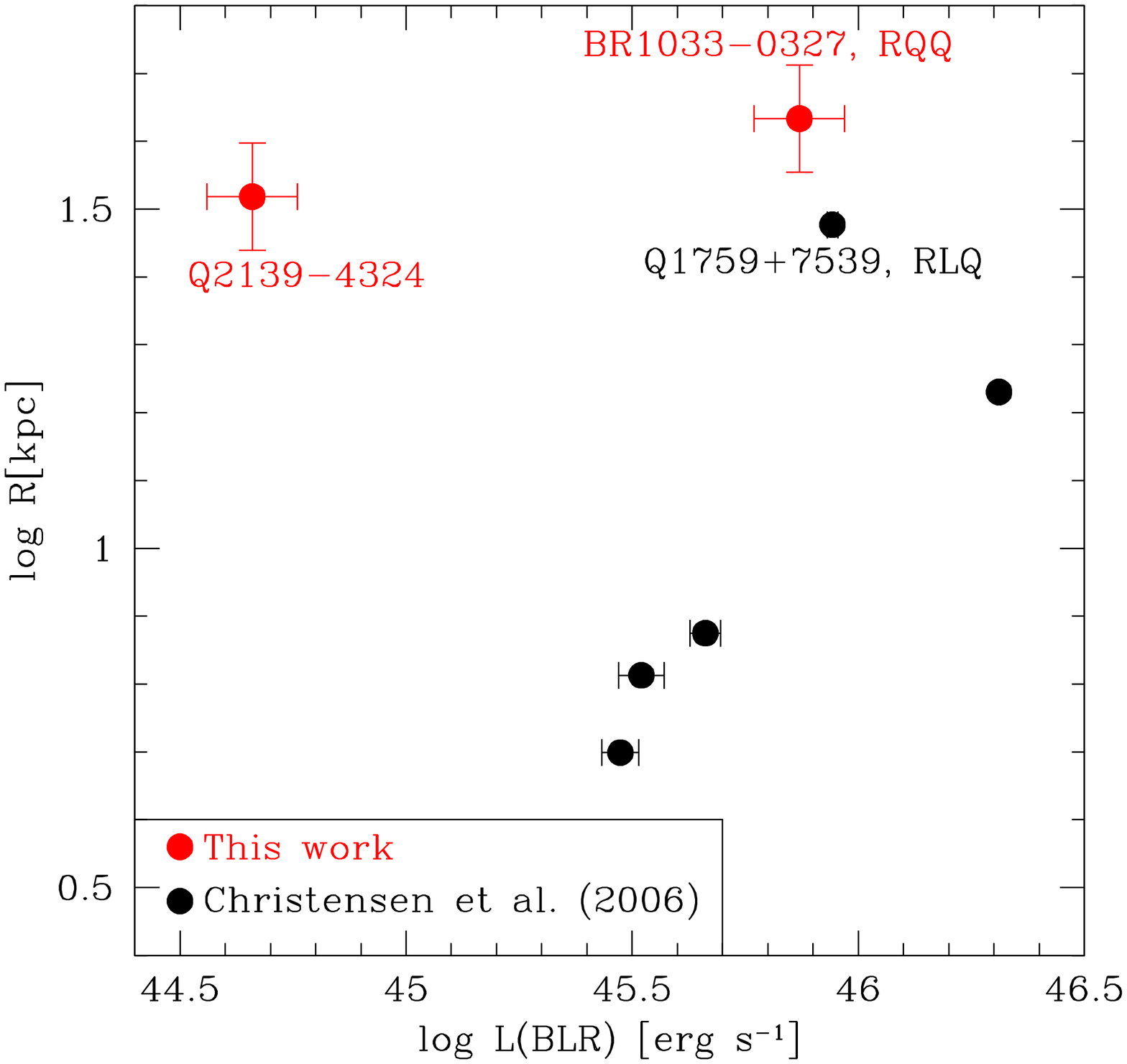}
\caption{{\it Top:}  relation between  the total luminosity  of 
the \lya\, envelopes  and that of the quasar  in the broad \lya\, line.
Our measurements  are compared with CJW.   The open  symbols represent
the direct  measurements  while the filled  symbols  are corrected for
slit clipping (see text).  The green symbols show our upper limit for
\sdss, where no \lya\, envelope is seen.  In doing this, we assume  
a size r=33 kpc (i.e., the mean size of the two  other objects) and we
do not correct  for slit-clipping.  {\it  Bottom}:  the radius of  the
envelope as a function of $L(BLR)$.  Note  the tight trend followed by
the envelopes of the bright RQQs of CJW, and the different behavior of
the envelopes of the fainter quasar \q. Note also that \br\, stands 
above the maximum size of the objects of CJW, likely due to our deeper
flux limit.}
\label{sb}
\end{figure}

\subsection{Redshift dependence}

The  observed surface brightness  of  the two \lya\, envelopes detected
here is fainter by about 1-2 orders of magnitude compared to the
CJW objects. This difference cannot be explained solely by
the different redshift ranges  of the two  samples. While our sample is
at    $z_1 \sim 4.5$,  most of the  objects   in  CJW are at
$z_2\sim 3.3$  and so the flux ratio by redshift dimming alone would
be   $F(z_2)/F(z_1) = (1+   z_2)^4   / (1+z_1)^4\sim 2.7$,  i.e., much
smaller than the observed ratio.

Taken  at  face value, our sample   as a  whole  finds that higher-$z$
objects tend to have smaller  surface brightnesses.  Nevertheless,  it
is  important to note  the possibly strong  dependence of the envelope
properties on the object luminosity  (see below). In particular,  only
one object  (\br) in  our  sample overlaps  with the  luminosity range
probed by CJW,  and its surface  brightness cannot be rejected at  the
$3-\sigma$ level from being drawn from the same sample (where $\sigma$
is the standard deviation and the  $t-$test performed on the logarithm
of  the quantities  rejects the object  at  the 94\%  level).  The two
other  objects  in our  sample  do  have  significantly  lower surface
brightness, at the 99.8\% confidence according to $t-$test.  This is a
lower limit given the upper limit on the undetected envelope in
\sdss.  However, the quasar luminosities are  also  lower by almost an
order of magnitude relative to the faintest object in CJW.

One potential  concern is  whether our observations  artificially find
larger envelopes   than CJW just   because they are deeper.   We note,
however, that the mean size  of the envelopes in  our sample is  $\sim
38$\,kpc compared  to $26.4$\,kpc in CJW's  sample, which could account
for a  50\%  difference in  area  but not  for  an  order of magnitude
effect.

\subsection{Dependence upon the luminosity of the quasar}

While  the surface  brightness of  the \lya\,  envelopes in  our small
sample of 3 objects is much lower than the  observations of CJW, their
total luminosities agree much better.

We show in Fig.~\ref{sb}  the luminosity-luminosity diagram, comparing
the total flux in the  \lya\, envelope as  a function of the  quasar's
luminosity in  the    broad   \lya\, line.   After    correcting   for
slit-clipping, we find that our points, combined with the ones of 
CJW, follow the linear relation:

\begin{equation}
{\rm Log} [L(Ly\alpha)] = 1.8 \times {\rm Log}[L({\rm BLR})] - 39.2
\end{equation}

This fit considers a mix of objects at  redshifts $z\sim 3$ and $z\sim
4.5$,  i.e., we assume no strong  redshift evolution of the luminosity
of the envelopes.  We do  not include in the fit  our upper limit  for
\sdss although the data point is shown in  Fig.~\ref{sb}. The
depth of our observations for this object is consistent with the trend
found by using the 5 points  of CJW and our 2  objects with a positive
detection  of  a \lya\,   envelope.   Our new  observations  for the 3
objects therefore seem to support the fact  that brighter quasars also
have  brighter  \lya\, envelopes,  under  the assumption of negligible
redshift evolution.

The  relation between   the   size of the   envelope  and  the  quasar
luminosity is  much less clear.  CJW find  a trend  that brighter RQQs
also display larger  envelopes.  In our sample  we find no such trend,
as  shown  in  Fig.~\ref{sb}.    The  main  difference   between   the
observations of CJW and ours  is depth.  With our deeper observations,
we probe  \lya\,  envelopes much  further away   from the quasar  than
CJW. In addition  the small field of view  used in CJW  implies severe
clipping  of  the   envelopes,  if they   extend  much   beyond  a few
arcsecs.  This  may also be  at the  origin of  the discrepant surface
brightnesses between CJW and the present work.

\section{Comparison with models and future prospects}

Current models for  the emission  of quasar   envelopes are in   their
infancy and  estimates for the  observable properties  of such objects
depend on several  tunable  parameters such   as halo  mass,  clumping
factor for the gas, quasar luminosity, metallicity, as well as redshift
\citep{HR01,AME02,CMBG08}. This  limitation, as well   as the size  of
current  samples and the  parameter range covered, make a quantitative
comparison with current models  unwarranted.  In addition, our results
may  be partly contaminated  by emission   from excited inter  stellar
medium in the host galaxy of  the quasar; an  effect not accounted for
by current models.  Nevertheless, several qualitative observations may
be made.

The flux level observed  from the envelopes  of bright quasars seem to
indicate that  such  objects inhabit halos which  are  more massive in
terms of their gaseous content (and possibly total mass) than those of
$L^\star$  galaxies \citep{AME02,CMBG08}.   Further  support for  this
conclusion comes from study of bright quasars at low redshifts
\citet{SE06}. The   new results presented here   for the less luminous
quasars indicate that     the surface brightness  is    lower than the
predictions of \citet{HR01} for  all relevant  halo masses (see  their
Fig. 2).  Within the  framework of their  model this suggests a  rapid
evolution of those systems already at $z> 4.5$, i.e. the cold gas has
been  converted to   stars (or into   a  thin disk) already at  $z\sim
4.5$. As a consequence,  there is not  enough gas left when the quasar
turns on,   to  reprocess the   quasar  light into  detectable  \lya\,
emission.

Our  results seem  to indicate  a much more   complex  picture than is
portrayed by current  models. To be able to  make some progress and to
quantitatively compare data and models it is  necessary to gain better
information concerning:

\begin{itemize}

\item The surface brightness profile of quasar envelopes. This will 
allow to make a  better  distinction  between recombination flux   and
photon scattering models  as well as to  understand the  origin of the
gas and  possible  link with absorption-line  systems  associated with
galaxies \citep{AME02,CMBG08}.

\item The  gas metallicity (via the  detection of  e.g., \ion{N}{V}\,$\lambda
1240$ line). This would provide an essential clue to the origin of the
gas  and  whether or not   it is considerably  enriched.   It may also
reveal whether the cooling time-scales  and  gas densities assumed  in
different models are realistic.

\item Dependence of envelope   observables  on quasar properties.  This  will
allow us to discriminate between the effect of environment and that of
quasar  luminosity on the properties  of the envelope thereby allowing
us to constrain its physics.

\end{itemize}

So far, the observational material available  to study \lya\, envelopes
of RQQs  is extremely  scarce, with  a  mere  half-dozen objects  with
different redshifts and observed with  different instruments, down  to
very different depths.  

Our deep slit-spectroscopy observations show that a homogeneous sample
of quasars with redshift  up to  z=4.5 can  be  built in a  reasonable
amount   of telescope time.   In   addition, our  finding that  \lya\,
envelopes can be  surprisingly large (r$\sim$10-15\arcsec)  and faint,
make  integral field  spectroscopy a  poor  solution to undertake this
task, due to poor cosmetics and sensitivity.  A  more viable option is
to conduct a two-step programme  focusing  on 1- narrow-band imaging  of
RQQs at different redshift slices  in order to map surface  brightness
profiles  for the   \lya\, envelopes of   quasars with  a  broad range
luminosities  and 2- to  carry  out  deep slit-spectroscopy  with well
controled  slit-clipping, with high spectral  resolution and with very
deep  flux  limit. The latter  shall  provide  us with  the additional
metallicity and velocity information required to constrain models.

\begin{acknowledgements}
We would like to thank Dr. Lise  Christensen for providing us with the
electronic  form of  the Tables in  CJW  and Anne  Verhamme for useful
discussions.  This study  is supported by  the Swiss National  Science
Foundation.
\end{acknowledgements}

\bibliographystyle{aa}
\bibliography{agn}

\end{document}